\documentclass[preprint,12pt]{elsarticle}%
\usepackage{amssymb}
\usepackage{amsfonts}
\usepackage{amsmath}
\usepackage{graphicx}%
\providecommand{\U}[1]{\protect\rule{.1in}{.1in}}
\graphicspath{{E:/dottorato/Coulomb_damper/Testo/Paper/Immagini/}}
\journal{journal}

\begin{document}
%
\begin{frontmatter}%


%

\title{An upper bound to multiscale roughness-induced adhesion enhancement}%

%

\author{M. Ciavarella}%
%

\address
{Politecnico di BARI. Center of Excellence in Computational Mechanics. Viale Gentile 182, 70126 Bari. Mciava@poliba.it}%
%

\begin{abstract}%

Recently Guduru and coworkers have demonstrated with neat theory and
experiments that both increase of strength and of toughness are possible in
the contact of a rigid sphere with concentric single scale of waviness,
against a very soft material. The present note tries to answer the question of
a multiscale enhancement of adhesion, considering a Weierstrass series to
represent the multiscale roughness, and analytical results only are used. It
is concluded that the enhancement is bounded for low fractal dimensions but it
can happen, and possibly to very high values, whereas it is even unbounded for
high fractal dimensions, but it is also much less likely to occur, because of
separated contacts.%

\end{abstract}%
%

%

\end{frontmatter}%



\section{\bigskip Introduction}

Guduru and collaborators (Guduru (2007), Guduru \&\ Bull (2007), Waters
\textit{et al} (2009)) have recently considered a model in which a sphere has
a superposed waviness, as defined by the axisymmetric form%
\begin{equation}
f\left(  r\right)  =\frac{r^{2}}{2R}+A\left(  1-\cos\frac{2\pi r}{\lambda
}\right)
\end{equation}
i.e. with concentric waviness, where $R$ is the sphere radius, $\lambda$ is
wavelength of roughness (see an example in Fig.1). Guduru also shows that
similar results are obtained if a plane roughness is assumed, similar to the
function above by with $x-$coordinate rather than $r$. Guduru shows that very
significant (one order of magnitude) increase of strength as well as toughness
can be obtained by adding roughness, i.e. with respect to the smooth case. It
should be immediately remarked that Jin et al (2011) have since then shown
that some of the enhancement obtained by Guduru is specific to this assumption
(either axisymmetric or purely 1D roughness), and therefore we may expect much
less enhancement for, say, random roughness.

\begin{center}
$%
\begin{array}
[c]{cc}%
{\includegraphics[
height=2.6072in,
width=3.6611in
]%
{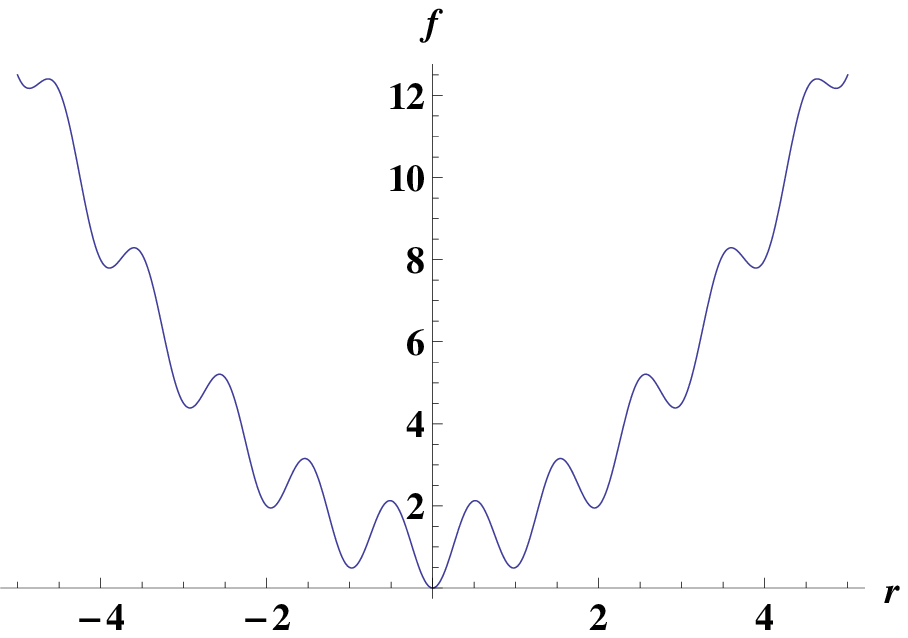}%
}
&
\end{array}
$

Fig.1 The Guduru sphere for $R=\lambda=A=1$
\end{center}

The concentric waviness permits a quite simple exact axisymmetric analysis,
assuming a simply connected contact area develops. Already for a single
waviness as in Guduru (2007), there are some limitations for this solution to
hold, as clearly for "sufficiently" large amplitude of roughness a realistic
solution will show some separated contacts. Also, Waters \textit{et al} (2009)
have clarified that much of the enhancement comes from the assumption of JKR
regime, and therefore one needs to check also the "Tabor parameter".

We shall here try to repeat some of the Guduru (2007) aspects of the solution,
in the context of a multiscale roughness, as it is more likely to occur in
practical cases, using for simplicity a Weierstrass series instead of a single
sinusoid, which was used in related contexts in Ciavarella et al (2000)
without adhesion for the fully separated regime, and by Afferrante et al
(2015) with adhesion, but with limited results concerning loading phase.
Specifically, we assume%
\begin{equation}
f\left(  r\right)  =f_{0}\left(  r\right)  +\mathfrak{g}_{0}\sum_{n=0}%
^{\infty}\gamma^{\left(  D-2\right)  n}\cos\left(  2\pi\gamma^{n}r/\lambda
_{0}\right)  \label{z_W}%
\end{equation}
where $f_{0}\left(  r\right)  $ is a "smooth profile" defining function, which
is a convex punch for example $f_{0}\left(  r\right)  =\frac{r^{2}}{2R}$ -- we
introduce this to avoid having to deal with a fully periodic surface, for
which the "smooth" behaviour is itself more difficult to define. If $\gamma>1$
and $D>1,$eq. (\ref{z_W}) defines, in a plane section, a plane fractal surface
of fractal dimension $D\ $(the real surface dimension will be one unit
higher), where we have
\begin{equation}
\mathfrak{g}_{n}=\mathfrak{g}_{0}\gamma^{\left(  D-2\right)  n},\qquad
\lambda_{n}=\lambda_{0}\gamma^{-n}%
\end{equation}
and hence the radius at given scale $n$ is $R_{n}=\frac{1}{\mathfrak{g}_{n}%
}\left(  \frac{\lambda_{n}}{2\pi}\right)  ^{2}=\frac{1}{\mathfrak{g}_{0}}%
\frac{\lambda_{0}^{2}}{4\pi^{2}}\gamma^{-Dn}$.

Fig.2 plots some examples of rough spheres so produced. Notice that the
roughness may equally be present in the other body, although Guduru for his
experiments considered a rough rigid sphere against an elastic nominally flat material.

\begin{center}
$%
\begin{array}
[c]{cc}%
{\includegraphics[
height=2.4686in,
width=3.6611in
]%
{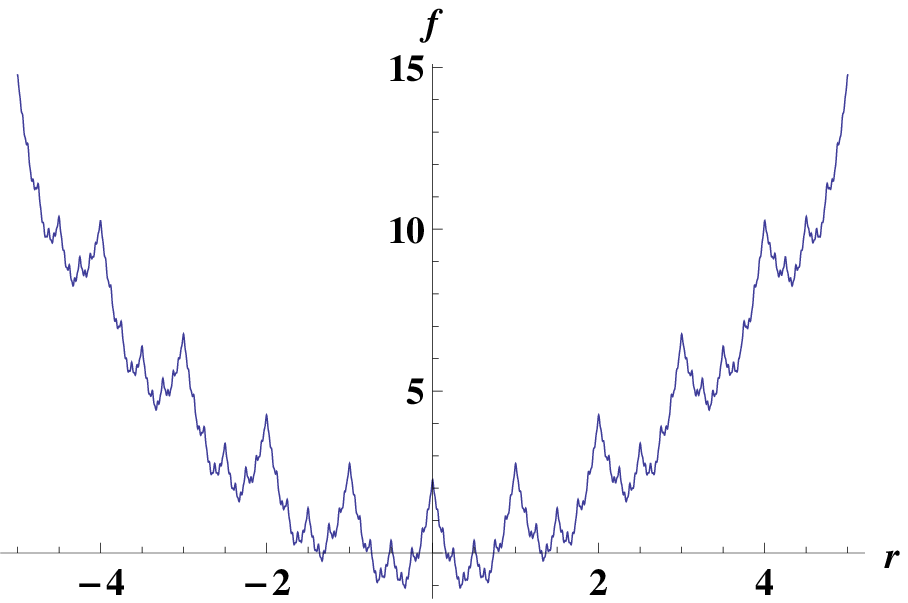}%
}
& (a)\\%
{\includegraphics[
height=2.4541in,
width=3.6611in
]%
{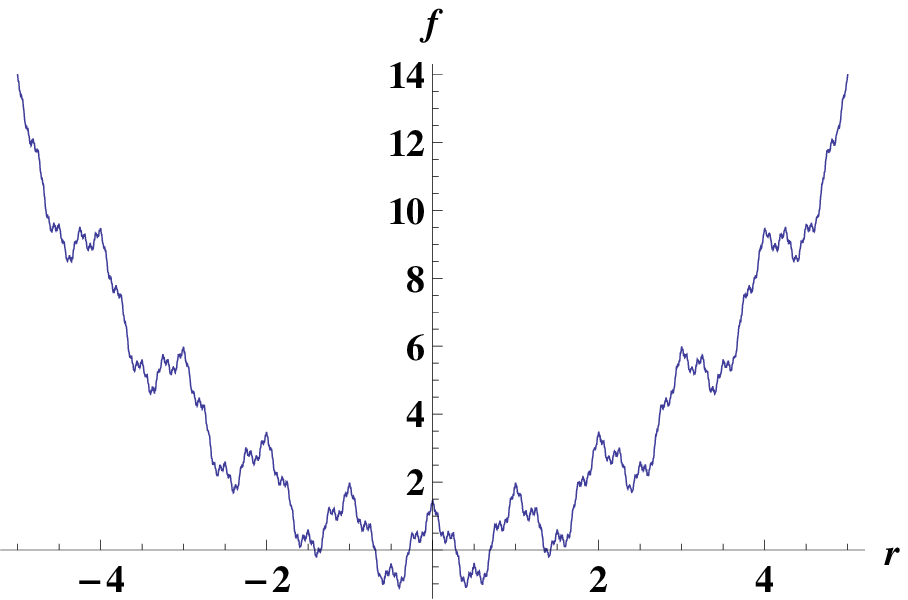}%
}
& (b)
\end{array}
$

Fig.1 The Weierstrass sphere for $R=g_{0}=\lambda=1$, $D=1.2$. For (a)
$\gamma=2$ and (b) $\gamma=4$
\end{center}

\section{\bigskip Some results}

\bigskip Waters \textit{et al} (2009) give a good summary of Guduru's theory
and experiments: it is shown that the load oscillates when it crosses a crest
of a wave, and this results in highly "wavy" curves. We will not give detailed
account of this theory, as we shall instead concentrate on an asymptotic
expansion solution (which permits, by joining all the minima and maxima of the
resulting function, also to obtain an "envelope" solution) given by Kesari et
al (2011) for small wavelength, in particular $\lambda<<a$, where $a$ is the
contact area radius.

Kesari et al (2011) suggest that if roughness is described by a function
$\lambda_{0}\varrho\left(  r/\lambda_{0}\right)  ,$ where the dimensionless
function $\varrho\left(  r/\lambda_{0}\right)  $ can be expanded in Fourier
series. Here, we shall use directly the Kesari result as a special case for
the Weierstrass series in order to get deterministic results for the maxima
and minima. Weierstrass is in fact a restricted form of Fourier series as we
shall consider $\gamma$ as integer and%
\begin{equation}
\varrho\left(  \xi\right)  =\sum_{n=0}^{\infty}a_{n}\cos\left(  2\pi\gamma
^{n}\xi\right)
\end{equation}

According to the Kesari et al (2011) expansion, the equilibrium curves are
described by load $P_{K}\left(  a\right)  $ and approach $h_{K}\left(
a\right)  $%
\begin{align}
P_{K}\left(  a\right)   &  =P_{M}\left(  a\right)  -E^{\ast}\sqrt{2\pi
a^{3}\lambda_{0}}\rho\left(  a/\lambda_{0}\right) \\
h_{K}\left(  a\right)   &  =h_{M}\left(  a\right)  -\sqrt{\frac{\pi a\lambda
}{2}}\rho\left(  a/\lambda_{0}\right)
\end{align}
where $E^{\ast}$ is plane strain elastic modulus, $h_{M}\left(  a\right)
,P_{M}\left(  a\right)  $ correspond to the smooth profile solution, and for
$\xi=a/\lambda_{0}$, the function $\rho\left(  \xi\right)  $ is given by%
\begin{equation}
\rho\left(  \xi\right)  =\sum_{n=0}^{\infty}\sqrt{2\pi\gamma^{n}}\left[
-a_{n}\sin\left(  2\pi\gamma^{n}\xi-\frac{\pi}{4}\right)  \right]
\end{equation}

Guduru's case is recovered when $a_{0}=A/\lambda=A/\lambda_{0}$, and the
macroscopic shape $f_{0}\left(  r\right)  $ is Hertzian parabola. To find the
envelope, one simply needs to take the maxima and minima of the equilibrium
curve, which are trivial for a single sinusoid. In fact, in this case%
\begin{align}
P_{K}\left(  a\right)   &  =P_{M}\left(  a\right)  \pm2\pi E^{\ast}\frac
{A}{\lambda_{0}}\sqrt{a^{3}\lambda_{0}}\\
h_{K}\left(  a\right)   &  =h_{M}\left(  a\right)  \pm\pi\frac{A}{\lambda_{0}%
}\sqrt{a\lambda_{0}}%
\end{align}

Before proceeding further, let us notice that an interesting feature emerges
in general, and that is that the smooth profile solution $h_{M}\left(
a\right)  ,P_{M}\left(  a\right)  $ contains a profile-independent
contribution (which essentially is the flat punch solution term in the JKR
process), and a profile dependent part $h_{M,profile}\left(  a\right)
,P_{M,profile}\left(  a\right)  $.\ With this separation, for example using
2.12a, 2.13a of Kesari et al (2011), one can derive at the quite general
expressions for the Weierstrass series roughness%
\begin{align}
P_{K}\left(  a\right)   &  =P_{M,profile}\left(  a\right)  -a^{3/2}\sqrt{8\pi
wE^{\ast}}\left(  1\pm\frac{1}{\alpha_{0}\sqrt{\pi}}\sum_{n=0}^{\infty}%
\sqrt{\gamma^{n}+1}\left[  \gamma^{\left(  D-2\right)  n}\sin\left(
2\pi\gamma^{n}\xi-\frac{\pi}{4}\right)  \right]  \right) \\
h_{K}\left(  a\right)   &  =h_{M,profile}\left(  a\right)  -a^{1/2}\sqrt
{\frac{2\pi w}{E^{\ast}}}\left(  1\pm\frac{1}{\alpha_{0}\sqrt{\pi}}\sum
_{n=1}^{\infty}\sqrt{\gamma^{n}+1}\left[  \gamma^{\left(  D-2\right)  n}%
\sin\left(  2\pi\gamma^{n}\xi-\frac{\pi}{4}\right)  \right]  \right)
\end{align}
where
\begin{equation}
\alpha_{0}=\sqrt{\frac{2w\lambda_{0}}{\pi^{2}E^{\ast}g_{0}^{2}}}%
\end{equation}
is the parameter Johnson (1995) introduced for the JKR adhesion problem of a
nominally flat contact with a single scale sinusoidal waviness of amplitude
$g_{0}$ and wavelength $\lambda_{0}$.

In the general case, if we had used a Fourier representation of roughness, we
would not have known how the maxima and minima of the various Fourier
components could combine. But as here we are considering a Weierstrass series
and we can take $\gamma>>1$, then the maxima and minima simply sum
algebraically, leading to the envelope (assuming $\sqrt{\gamma^{n}+1}%
\simeq\gamma^{n/2}$)%
\begin{align}
P_{K,env}\left(  a\right)   &  =P_{M,profile}\left(  a\right)  -a^{3/2}%
\sqrt{8\pi wE^{\ast}}\left(  1\pm\frac{1}{\sqrt{\pi}}\sum_{n=0}^{\infty}%
\frac{1}{\alpha_{n}}\right) \\
h_{K,env}\left(  a\right)   &  =h_{M,profile}\left(  a\right)  -a^{1/2}%
\sqrt{\frac{2\pi w}{E^{\ast}}}\left(  1\pm\frac{1}{\sqrt{\pi}}\sum
_{n=0}^{\infty}\frac{1}{\alpha_{n}}\right)
\end{align}
where we introduced a scale-dependent Johnson parameter%
\begin{equation}
\alpha_{n}=\alpha_{0}\left(  \gamma^{\left(  2D-3\right)  n}\right)
^{-1/2}=\frac{\alpha_{0}}{\gamma^{\left(  D-3/2\right)  n}} \label{alfan}%
\end{equation}

The series defined by the sum of the Johnson parameters converges for all
$D<1.5$ as
\begin{equation}
\lim_{N\rightarrow\infty}\sum_{n=0}^{N}\frac{1}{\alpha_{n}}=\frac{1}%
{\alpha_{0}}\lim_{N\rightarrow\infty}\sum_{n=0}^{N}\gamma^{\left(
D-3/2\right)  n}=\frac{1}{\alpha_{0}}\lim_{N\rightarrow\infty}\left(
\frac{\gamma^{\left(  D-3/2\right)  \left(  1+N\right)  }-1}{\gamma^{\left(
D-3/2\right)  }-1}\right)  =\frac{1}{\alpha_{0}}\left(  \frac{1}%
{1-\gamma^{\left(  D-3/2\right)  }}\right)
\end{equation}
which is the case of common interest for fractal surfaces, and which is the
case where we can expect more easily enhancement anyway since alternative
solutions during loading phase only (Afferrante et al, 2015) show that only in
this case of $D<1.5$ we expect a limit contact area due to infinite roughness:
for higher fractal dimension, the contact resembles increasingly that obtained
in the absence of adhesion, as in the classical (Ciavarella et al., 2000)
solution for the Weierstrass profile.

With respect to the smooth surface therefore, it is easy to show that we have
obtained the amplification factor for pull off as
\begin{equation}
F\left(  \alpha_{0},\gamma,D\right)  =\left(  1+\frac{1}{\alpha_{0}\sqrt{\pi}%
}\left(  \frac{1}{1-\gamma^{\left(  D-3/2\right)  }}\right)  \right)  ^{2}%
\end{equation}
which is plotted in Fig.3 for representative values.

\bigskip

\begin{center}
$%
\begin{array}
[c]{cc}%
{\includegraphics[
height=2.4966in,
width=3.6611in
]%
{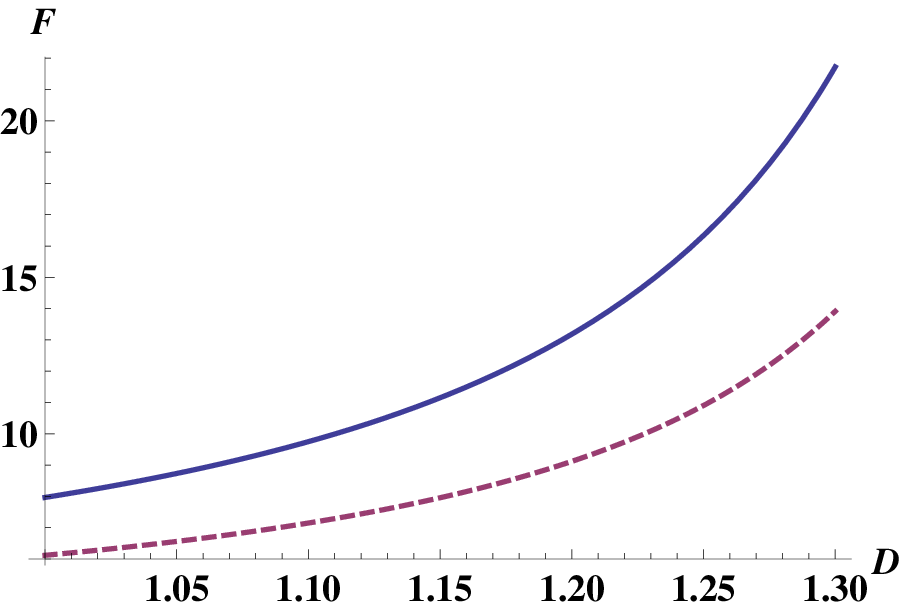}%
}
& (a)\\%
{\includegraphics[
height=2.4966in,
width=3.6611in
]%
{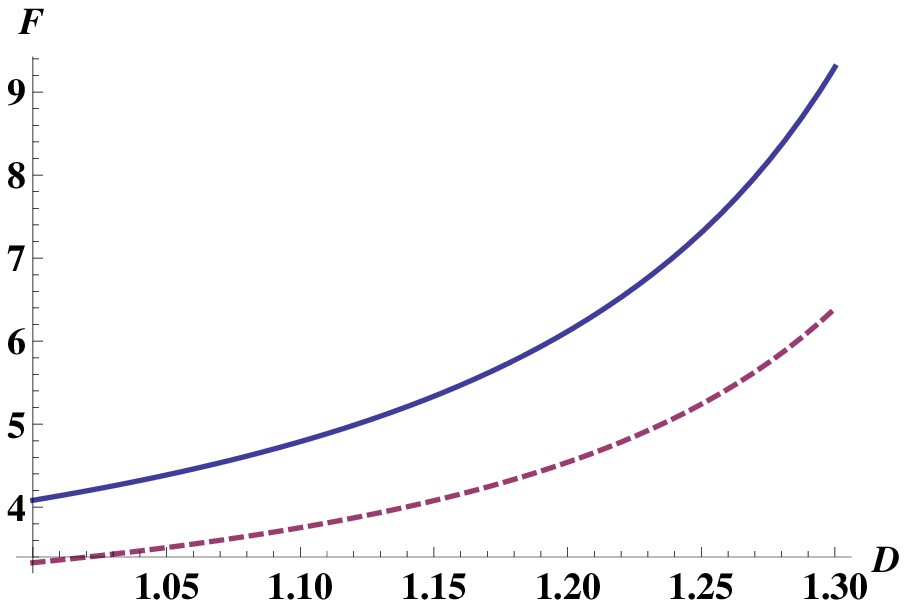}%
}
& (b)
\end{array}
$

Fig.3 Pull-off amplification factor $F\left(  \alpha_{0},\gamma,D\right)  $
for $\alpha_{0}=0.56$ (a) and $\alpha_{0}=1$ (b). In both cases, thick solid
line is $\gamma=5$, and dashed line is $\gamma=10$.
\end{center}

We have taken in Fig.3 values of $\gamma$ relatively high, as otherwise there
is no guarantee that our procedure in estimating the cumulative scale effect
was accurate. Therefore, we chose $\gamma=5$ and $\gamma=10$ while Fig.3a and
Fig.3b distinguish for the initial value of the Johnson's parameter. As it is
evident also from Fig.3, this factor can be extremely high for the fractal
dimensions near $D=1.5$, and this is not so surprising considering that we
found the amplification doesn't converge for higher $D$. Notice also that the
for the chosen parameters of $\alpha_{0}=0.56,1$, the amplification for a
single wave of roughness were respectively 4 and $\left(  1+\frac{1}{\sqrt
{\pi}}\right)  ^{2}=\allowbreak2.45$, and therefore the multiscale additional
enhancement is really significant.

Obviously similar remarks can be made regarding the factor $\alpha_{0}$: the
lower this factor (and therefore the lower the adhesion effect in the Johnson
solution of a single sinusoid), the higher the amplification. So it would seem
from this analysis that the highest amplifications would occur for low
$\alpha_{0}$ and high fractal dimensions, which is exactly the case where we
expect more likely separated contact! This is simply an indication that
separation is more and more likely the higher the amplification is expected to
be in the simply connected contact area model we are assuming.

\section{Discussion}

There are various reasons why the amplifications in adhesion predicted by the
theory are limited. We shall discuss them separately in order.

\subsection{JKR assumption}

JKR theory is strictly valid, in the classical JKR case of a sphere, when
$\mu>5$, but JKR works well for $\mu>0.3$ in practice: below $\mu=0.3$, the
behaviour approaches that of a rigid sphere. In the original case of the
sphere, JKR and rigid theory only differ by a small prefactor in the pull-off
loads, but Tabor parameter was shown in Waters et al (2009) to limit
enhancement of the Guduru problem considerably, namely $\mu<1$ at the scale of
the sphere completely destroys the enhancement. Following Waters et al (2009),
we can define, for multiscale roughness, a scale-dependent Tabor parameter
\begin{equation}
\mu_{n}=\frac{\sigma_{th}}{E^{\ast}}\left(  \frac{9}{2\pi}\frac{R_{n}}{l_{a}%
}\right)  ^{1/3}=\mu_{0}\gamma^{-Dn/3}%
\end{equation}
where $\sigma_{th}$ is theoretical strength of the material, $E^{\ast}$
elastic modulus, $l_{a}=w/E^{\ast}$ and $R_{n}$ which is not a radius of a
sphere, nevertheless can serve to estimate the role of elastic deformation at
the $n-th$ scale. It is clear that, even if at scale 0 the Tabor parameter is
well in the JKR regime, with finer scales of roughness, the Tabor parameter
would tend to reduce quite fast. For example, with $\gamma=10$ and even with
low $D=1.2$, this reduces Tabor parameter at the macroscale to 1/10 just with
one additional scale of roughness. Therefore, this factor alone will limit
very substantially practical evidence of multiscale enhancement.

\subsection{Loading dependence}

In the original Guduru (2007) problem of a single sinusoid, the condition
$\alpha_{0}>0.56$ was seen to correspond to self-flattening of waviness,
irrespective of the applied load, as in that case it was shown to correspond
also to imposing monotonicity of the profile, so that the solution was
obtained without need of a proper, sufficiently high loading stage. Hence, the
amplifications with this range was likely to occur. Obviously, this is the
range where a single waviness, even from the equations above, amplification is
lower than 4. But for multiscale roughness, we could increase this.
Unfortunately, we do not know if $\alpha_{0}>0.56$ guarantees that
self-flattening occurs on all scales. A first consideration seems to suggest
that, for $D<1.5$, as $\alpha_{n}$ increase, a fortiori there should be
self-flattening on all scales. However, we should check if there is also
monotonicity of the profile --- as otherwise, even if $\alpha_{n}>0.56$,
separated points in the profile may simply not be in a condition to jump into
contact. Hence, it is useful therefore to extend Guduru's analysis of
monotonicity to Weierstrass, by taking the derivative of the surface function%
\begin{equation}
z^{\prime}\left(  r\right)  =f_{0}^{\prime}\left(  r\right)  +\mathfrak{g}%
_{0}\frac{2\pi}{\lambda_{0}}\sum_{n=0}^{\infty}\gamma^{\left(  D-1\right)
n}\sin\left(  2\pi\gamma^{n}r/\lambda_{0}\right)
\end{equation}

The second term of this is obviously is related to the full contact pressure
under adhesionless conditions of the Weierstrass profile contact (Ciavarella
et al., 2000)
\begin{equation}
p\left(  x\right)  =\bar{p}+\sum_{n=0}^{\infty}p_{n}^{\ast}\cos\left(
2\pi\gamma^{n}x/\lambda_{0}\right)  ,\qquad\bar{p}-\hat{p}\leq p\left(
x\right)  \leq\bar{p}+\hat{p}%
\end{equation}
being%
\begin{align}
p_{n}^{\ast}  &  =\pi E^{\ast}\mathfrak{g}_{0}\gamma^{\left(  D-1\right)
n}/\lambda_{0}=p_{0}^{\ast}\gamma^{\left(  D-1\right)  n}\\
\hat{p}  &  =\sum_{n=0}^{\infty}p_{n}^{\ast}=p_{0}^{\ast}\sum_{n=0}^{\infty
}\gamma^{\left(  D-1\right)  n} \label{ps_n}%
\end{align}

For $\gamma>1$ and $D>1$, the series (\ref{ps_n}) does not
converge\footnote{Indicating that there is no finite value of mean pressure
$\bar{p}$ that is sufficient to ensure complete contact between a fractal
rigid surface of the Weierstrass form and an elastic half-plane in the
adhesiveless case.}. This suggests that the monotonicity condition which
Guduru could guarantee for just one term of waviness, independently on
loading, becomes increasingly more difficult to satisfy. Already with 2
scales, we find that one needs to rely on the loading process to find
"complete" (simply connected) contact over the contact area. The amplification
factor above could however be considered an upper bound, which could be
reached upon sufficient pre-loading. Adhesion will certainly show pressure-sensitivity.

\subsection{Kesari's envelope validity}

We have obtained the amplification factors under the implicit assumptions that
Kesari's envelope works. In the single scale waviness of Guduru, this was true
for \textit{low values of }$\beta_{G}=\frac{\lambda^{3}E^{\ast}}{2\pi wR^{2}}%
$. Here, we can define a scale-dependent $\beta_{n}$ looking at the scale $n$
and the waviness at scale $n+1$,
\begin{equation}
\beta_{n}=\frac{\lambda_{n+1}^{3}E^{\ast}}{2\pi wR_{n}^{2}}=\frac{16\pi
^{4}E^{\ast}}{2\pi w}\frac{\mathfrak{g}_{0}^{2}}{\lambda_{0}}\gamma
^{-3n-3+2Dn}%
\end{equation}
which rapidly goes to zero \textit{only for low fractal dimensions}, and this
suggests the Kesari envelope is increasingly more appropriate for this, most
important case. Therefore, at least this assumption is not particularly restrictive.

\section{Conclusions}

We have attempted to extend the Guduru model to multiscale roughness, using a
Weierstrass series. We found some estimates of the potential amplification,
which is higher than that of the single scale of waviness. We find in
particular that the potential amplification is bounded for $D<1.5$ and is
unbounded otherwise. However, many limitations suggest this amplification is
often impractical to reach: the assumption of JKR regime becomes increasingly
invalid for finer scales, the monotonicity of the profile, needed to guarantee
simply connected contact area, is also very unpractical to reach, and the
highest amplifications occur exactly where the assumption of a simply
connected area is most difficult to satisfy. Finally, a true 1D or
axisymmetric roughness is less common than random roughness, although one can
contrive some systems to wrinkle only in one direction and therefore it is not
unconceivable. When the roughness is random, this further reduces the expected amplifications.

\section{References}

Afferrante, L., Ciavarella, M., \& Demelio, G. (2015). Adhesive contact of the
Weierstrass profile. In Proc. R. Soc. A (Vol. 471, No. 2182, p. 20150248). The
Royal Society.

Ciavarella, M., Demelio, G., Barber, J. R., \& Jang, Y. H. (2000). Linear
elastic contact of the Weierstrass profile. In Proceedings of the Royal
Society of London A: Mathematical, Physical and Engineering Sciences (Vol.
456, No. 1994, pp. 387-405). The Royal Society.

\bigskip Guduru, P.R. (2007). Detachment of a rigid solid from an elastic wavy
surface: theory J. Mech. Phys. Solids, 55, 473--488

Guduru, P.R. , Bull, C. (2007). Detachment of a rigid solid from an elastic
wavy surface: experiments J. Mech. Phys. Solids, 55, 473--488

Jin, C., Khare, K., Vajpayee, S., Yang, S., Jagota, A., \& Hui, C. Y. (2011).
Adhesive contact between a rippled elastic surface and a rigid spherical
indenter: from partial to full contact. Soft Matter, 7(22), 10728-10736.

Johnson, K. L., K. Kendall, and A. D. Roberts. (1971). Surface energy and the
contact of elastic solids. Proc Royal Soc London A: 324. 1558.

Kesari, H., Doll, J. C., Pruitt, B. L., Cai, W., \& Lew, A. J. (2010). Role of
surface roughness in hysteresis during adhesive elastic contact. Philosophical
Magazine \& Philosophical Magazine Letters, 90(12), 891-902.

Kesari, H., \& Lew, A. J. (2011). Effective macroscopic adhesive contact
behavior induced by small surface roughness. Journal of the Mechanics and
Physics of Solids, 59(12), 2488-2510.

Waters, J.F. Leeb, S. Guduru, P.R. (2009). Mechanics of axisymmetric wavy
surface adhesion: JKR--DMT transition solution, Int J of Solids and Struct 46
5, 1033--1042

\end{document}